# Nonlinear Integrated Microwave Photonics

David Marpaung* and Benjamin J. Eggleton
Centre for Ultrahigh bandwidth Devices for Optical Systems (CUDOS), the Institute of Photonics and Optical Sciences (IPOS),
School of Physics, University of Sydney, NSW 2006, Australia
*e-mail: d.marpaung@physics.usyd.edu.au

*Abstract*— Harnessing nonlinear optical effects in a photonic chip scale has been proven useful for a number of key applications in optical communications. Microwave photonics can also benefit from the adoption of such a technology, creating a new concept of nonlinear integrated microwave photonics. Here, we discuss the potential of on-chip nonlinear processing towards the creation of robust and multifunctional microwave photonic (MWP) processors. We also highlight key recent results in the field, including frequency agile MWP filters and ultra-wideband signal generators.

## I. Introduction

Recently, there is a significant interest in exploiting photonic integrated circuits (PICs) for the processing of microwave signals [1]. The term integrated microwave photonics (IMWP) is used to describe research activities in this area. IMWP clearly promises a number of advantages compared to conventional fiber-based approach. The potential of integrating a number of different signal processing functionalities in a single chip will lead to significant reduction in size, weight, power consumption, and cost. Moreover, the ability to integrate various functions like modulation, passive signal processing, and detection in a single chip will lead to reduction in insertion losses, which is the key for MWP signal processor to compete with RF signal processor in terms of performance.

Thus far, IMWP has largely been limited to linear optical processing. It is interesting, however, to examine the potential of nonlinear optics in integrated platform for microwave signal processing. Optical nonlinearities (e.g., $\chi^{(2)}$ and $\chi^{(3)}$) [2], such as sum and difference-frequency generation (SFG/DFG), cross-phase modulation (XPM), degenerate/non-degenerate four-wave mixing (FWM), and stimulated Brillouin scattering (SBS), are suitable candidates for MWP signal processing. In the context of optical signal processing, these nonlinear optics effects have efficiently been harnessed, both in fibers and in integrated devices, for a number of functionalities such as all-optical delay [3], all-optical switching and multiplexing [4], and wavelength multicasting [5].

In this paper, we look at the potential of combining linear and nonlinear optical effects in a chip scale to enable a number of functionalities such as agile microwave filtering and spectrum channelization, with an ultimate goal to achieve a general purpose reconfigurable analog signal processor. We then review a number of recent IMWP demonstrations that exploited on-chip SBS, FWM, and XPM, in various platforms such as chalcogenide glasses and silicon. Finally, we discuss the prospects and as well as the challenges that should be tackled in order for nonlinear IMWP to thrive.

## II. Potential

The exponential growth in bandwidth demand of radio communications has put unprecedented challenges in the RF signal processing chain. Traditionally, strong emphasis in such a chain is put on digital signal processing (DSP). However, for increasing operating frequencies and larger bandwidths, a reliable analog signal processor (ASP) that can match the flexibility of a DSP is desired [6, 7]. Due to its high bandwidth and potential reconfigurability, IMWP processor is an attractive candidate for such a flexible ASP. However, current state-of the-art IMWP processors were developed only for very specific tasks, for example beamforming and filtering [8], frequency discrimination [9], or pulse shaping [10]. Thus, there is a lack of generality and multi-functionality of these processors. To emulate DSP, the IMWP processor has to be application agnostic, multi-purpose and programmable.

We believe that incorporating optical nonlinearities in an IMWP processor will open up the path towards a general purpose ASP. Gasulla and Capmany [11] have identified that a large number of MWP signal processing can be generalized as processing of a number copies of RF modulated optical signals using reconfigurable optical filters. Nonlinear optical processes that create new optical wavelengths such as FWM, SFG, or DFG are attractive candidates for generating such signal copies. In fact, this concept of wavelength multicasting has recently been implemented for MWP signal processing techniques such as channelization [12, 13] and reconfigurable filtering [14, 15]. In [12-14] FWM processes in highly nonlinear fibers (HNLF) were used, while SFG and DFG in PPLN have been exploited in [15]. In the view of the general processor, these wavelength copies will subsequently be processed by a reconfigurable optical filter. The typical signal processing tasks carried out by these filters include tunable time delay, carrier phase tuning, and complex (amplitude and phase) filtering. These processed optical signals will then be recombined or directed to separate dedicated outputs for a photodetection process.

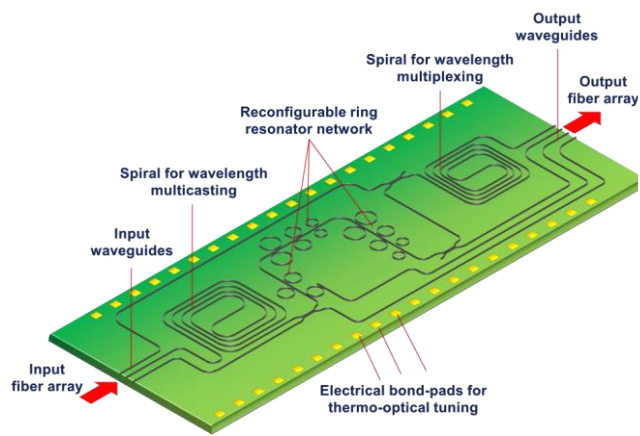

Fig. 1. (a) Illustration of the envisioned general purpose analog signal processor architecture.

The key challenge here is to create a monolithic photonic chip that host the nonlinear optics for wavelength multicasting and the reconfigurable linear optical filtering. Fig. 1 depicts an illustration of a possible design of such a photonic chip in $Si_3N_4$ technology. The chip consists of three main sections: two long spiral waveguides for wavelength multicasting and multiplexing based on cascaded FWM processes, and reconfigurable optical filter based on a network of ring resonators that can be tuned using thermo-optics effect. This ring network might host a mixture of all pass ring filters, add-drop rings, as well as Vernier (cascade of non-identical) rings which have been recently exploited for tunable delay lines in a multi-wavelength beamformer [16]. The programmability of the ring network will allow the chip to be reconfigured in real time, to synthesize different responses according to the user demand.

Obviously, other material platforms such silicon can also serve as good candidates to construct this ASP. However, $Si_3N_4$ waveguides might be more suited for this purpose due to the following characteristics: ultra-low propagation losses (approx. 0.1 dB/cm [17]) which is very important for MWP systems, moderately high optical nonlinearity (10 times of silica) [18-21] and free from nonlinear losses such as two-photon absorption (TPA) and free-carrier effect (FCA) [19]. On top of this, $Si_3N_4$ waveguides have shown compatibility with efficient thermo-optics tuning and is a CMOS compatible material. While silicon also tick some of the boxes and successful wavelength multicasting has been demonstrated in this platform [22, 23], the propagation loss of a typical SOI nanowires is relatively high (2-3 dB/cm) [24] and they suffer from TPA and FCA [4, 19]. These losses might be prohibitive for creating an efficient nonlinear IMWP processor.

### III. CURRENT STATUS

Very recently, a number of nonlinear IMWP demonstrations have been shown with impressive performance. For example, on chip SBS has been harnessed for creating an ultra-high performance tunable notch filter, while FWM in silicon nanowire has been used in the context of a reconfigurable multi-tap MWP filter. We believe this development is a significant first step towards the realization of a general purpose processor. Here, we review some these demonstrations.

#### A. *On-chip SBS filters*

Stimulated Brillouin scattering (SBS) is a nonlinear scattering process where a pump wave, when encounter acoustic vibrations in the medium, gets backscattered, resulting in Stokes and anti-Stokes waves. A probe, centered at the Stokes frequency, experiences gain when counter-propagated to the pump whereas a probe that is up-shifted with respect to the pump (anti-Stokes), experiences absorption. SBS has long been exploited in optical fibers for MWP signal processing [25, 26], but the recent demonstration of on-chip SBS [27] has finally enabled the control of this process in a compact photonic chip scale. The photonic chip was a 6.5-cm long chalcogenide glass (ChG) $As_2S_3$ optical waveguide with a cross-section of 4 μm × 850 nm and a large SBS gain coefficient ($g_B$ ~ 0.74 x $10^{-9}$ m/W). Using this photonic chip, key SBS RF signal processing such as tunable delay line [28] tunable bandpass [29] and notch [30, 31] filtering have recently been demonstrated.

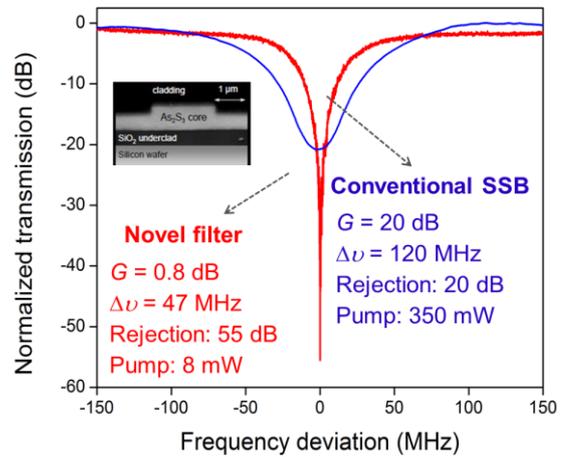

Fig. 2. Experimental result of MWP notch filter response using on-chip SBS with sub-1 dB gain. Blue trace indicates conventional filter using single sideband modulation and SBS loss spectrum [31]. Red trace indicates notch response obtained from novel complex sideband processing [34] and SBS gain spectrum. The novel technique leads to an increase in filter resolution (47 MHz full-width at half maximum) and ultra-high enhancement of the peak rejection (>55 dB), achieved with only 8 mW of pump power.

Recently, a remarkable progress has been made regarding MWP tunable notch filters based on SBS. A long standing problem in RF filters is to achieve wide frequency tuning simultaneously with high resolution filtering. For example, state-of-the-art absorptive band-stop filters are capable of a high peak attenuation (>50 dB) and high resolution (<10 MHz 3-dB isolation bandwidth measured from the passband) but have limited notch frequency tuning range, in the order of 1.4 GHz [32]. MWP notch filters, on the other hand, are capable of multi-gigahertz tuning, but are often limited in peak attenuation (< 40 dB) and resolution (gigahertz bandwidth instead of tens of megahertz). In [33], Marpaung et al. proposed a novel class of MWP notch filters with an ultrahigh peak rejection. In this novel scheme, an RF signal was encoded in the optical sidebands with unequal amplitudes and a phase difference using an electro-optic modulator (EOM). An optical filter (OF) was then used to equalize these sidebands amplitudes and to produce an anti-phase relation between them only in selected frequency region within the OF response. Upon photodetection, the beat signals generated from the mixing of the optical carrier and the two sidebands perfectly cancel at a specific microwave frequency, forming a notch with an anomalously high stopband rejection. Using SBS gain spectrum in 650 m of optical fibers as the OF, they realized an MWP notch filter with record performance (a high selectivity of ~10 MHz, an ultrahigh stopband rejection of >60 dB, wide continuous frequency tuning of 1-30 GHz, and flexible bandwidth reconfigurability 10-65 MHz). Very recently, this concept has also been implemented with SBS in compact ChG photonic chip. In [34], Marpaung et al. achieved a filter with an ultra-high suppression of more than 55 dB, a high resolution of 47 MHz, and ultra-wide tunable central frequency of 1-30 GHz. Remarkably, this was achieved using an ultra-low SBS gain of 0.8 dB, and a very low pump power of 7 mW. This represents 50 times reduction in required pump power, relative to a conventional SBS notch filter generated using single-sideband modulation scheme [31]. The measured responses of the novel and conventional filters are shown in Fig. 2. We believe that these results will ease the implementation of low power nanophotonic devices as high performance RF notch filters and will change the landscape of the field of integrated microwave photonics.

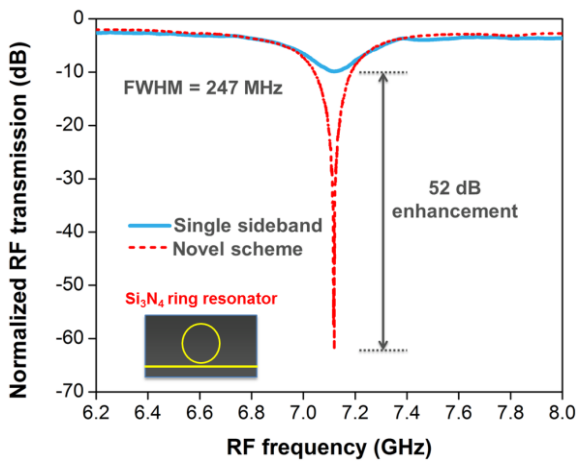

Fig. 3. Experimental result of MWP notch filter response using a low loss $Si_3N_4$ ring resonator [35]. Implementing the complex sideband processing [33-35] led to a peak rejection enhancement of > 50 dB relative to conventional single sideband scheme.

The key advantage of the novel notch filter scheme is that it is applicable for a wide range of optical filters. Marpaung et al. also implemented this technique using a reconfigurable silicon nitride ring resonator [35]. They achieved peak rejection enhancement of more than 50 dB (Fig. 3). Such a high suppression would normally occur only when the resonator is critically coupled. But the novel technique allows one to achieve this high contrast independent of the ring coupling, thereby significantly relaxing the fabrication tolerance of such ring resonators.

### B. On chip XPM and FWM

On-chip ultrafast Kerr nonlinearity has been exploited in the context MWP signal processing for ultra-wideband (UWB) signal generation [36] and reconfigurable multi-tap filtering [37]. In [36], Tan et al. combine the effects of cross-phase modulation (XPM) and birefringence in an $As_2S_3$ rib waveguide to generate polarity-inverted UWB monocycles with a single optical carrier. The high Kerr-nonlinearity of ChG in a chip platform enables efficient XPM in a short length of 7 cm. The combination of XPM and birefringence essentially acts as an all-optical differentiator that converts a train of Gaussian pulses into a train of monocycle electrical pulses after photo-detection process. Using this technique, Tan et al. demonstrated the generation of a wide variety of UWB pulses, including polarity-inverted monocycles and doublets. A key advantage in exploiting XPM was the chirp erasure of the input optical pulses, because XPM depends solely on the pulses intensity. Thus, shaped UWB optical pulses have a good tolerance to dispersion over fiber and are more suitable for long-distance transmission in UWB over fiber communication systems.

Very recently, Chen et al [37] have reported a step towards the integration of a reconfigurable MWP filter concept in [14], by exploiting an FWM process in a silicon nanowire, instead of using HLNF. The silicon waveguide ($h$= 220 nm, $w$= 650 nm, length= 12 mm) was dispersion engineered to yield broadband FWM. They generated 2 idlers with measured conversion efficiency of -25 dB. These wavelengths taps were then modulated and amplitude controlled using a waveshaper. The used 4 km of single mode fiber (SMF) as the dispersive medium required to generate time delay between the taps. The measured the response of a 4-tap MWP filter and demonstrated that the 12 mm silicon nanowire can be used to replace 1 km of silica HNLF without compromising the quality of the MPF response.

### IV. OUTLOOK

We believe that the growth of nonlinear IMWP will strongly depend on two factors: the efficiency of the associated on-chip nonlinear processes, and the integrability of the nonlinear platform of choice. Most of the previously reported results were demonstrated in a highly nonlinear chalcogenide glass. This platform enabled highly efficient nonlinear processes but it is relatively limited in terms integration potential. CMOS compatible platforms such as silicon or silicon nitride are more attractive due to the moderate to high nonlinearity but more importantly their potential of integrating linear and nonlinear optics components as well as modulators and detectors either monolithically, or through hybrid integration schemes. This also applies, for example, to the high performance notch filter concept reported in [33-35]. It is very attractive to implement such as scheme with on-chip SBS in silicon/silicon nitride hybrid optical structure as shown in [38].

In the view of generating the multipurpose ASP, we believe incorporating photonic crystal structures [39] will be highly attractive for a number of reasons. This structure can be made highly dispersive, to replace the optical fiber spool for the time delay element required in most signal processing tasks.

### V. CONCLUSIONS

In this paper we introduced an exciting new concept in microwave photonics, where on-chip linear and non-linear processes are combined to enable unprecedented RF signal processing capabilities. We also offered perspective on key factors that will determine the growth of this field.

ACKNOWLEDGMENT

D.M. would like to thank Prof. Jose Capmany and Maurizio Burla for useful discussions. The authors thank Blair Morrison, Ravi Pant, Duk-Yong Choi, Steve Madden, Barry Luther-Davies, Kang Tan, and Chris Roeloffzen for their contribution.